\documentclass[showpacs,amsmath,amssymb,prb]{revtex4}

\usepackage{graphicx}
\usepackage{dcolumn}
\usepackage{bm}

\begin{document}

\title{Chiral fluctuations in triangular antiferromagnets at $T \ll T_N$}

\author{A. V. Syromyatnikov}
 \email{syromyat@thd.pnpi.spb.ru}
\affiliation{Petersburg Nuclear Physics Institute, Gatchina, St.\ Petersburg 188300, Russia}

\date{\today}

\begin{abstract}

Chiral fluctuations in triangular antiferromagnets (TAFs) at $T \ll T_N$ are studied theoretically. The case of a ferromagnetic interaction along $c$ axis (which is directed perpendicular to the plane of the lattice), Dzyaloshinskii-Moriya interaction ${\bf D}\| c$ and a weak magnetic field ${\bf H}\| c$ is considered in detail. Previously, this model has been proposed to describe quantum TAF CsCuCl$_3$. Expressions for dynamical chirality (DC) are derived within the linear spin-wave approximation. In contrast to non-frustrated antiferromagnets, DC is found to be nonzero even at $D,H=0$ in a one-domain sample. We argue that this unusual behavior stems from the fact that a ground state of $XY$ and Heisenberg TAFs is characterized by an axial vector along which DC is directed.

\end{abstract}


\maketitle

\section{Introduction}

Spin chirality has attracted a lot of attention during last two decades (see Refs.~\cite{kawa,kawa2,mal,mal2,pl1,pl2,lee} and references therein). The main subjects of investigation were helical magnets and triangular antiferromagnets (TAFs). A new (chiral) universality class of phase transitions has been proposed for these systems by H. J. Kawamura (see Ref.~\cite{kawa} for a review). In particular, he observed that the dimension of their order parameter differs from those of antiferromagnets (AFs) on bipartite lattices. Let us discuss this point in detail for TAF.

Since an order parameter space is defined as a topological space isomorphic to the set of ordered states, \cite{mermin} we consider the properties of the ground state. Due to frustration it has $120^\circ$ spin structure. The ground state in this case can be characterized by chiralities of elementary triangles. The spin chirality is defined as \cite{kawa}
\begin{equation}
\label{c0}
{\bf C}_{ij} = [{\bf S}_i\times{\bf S}_j],
\end{equation}
where ${\bf S}_i$ is a spin on cite $i$. The spin chirality of each elementary triangle is defined as a sum of chiralities of its spins when bypassing the triangle clockwise. The following two variants are possible: angles between all neighboring spins are a) $120^\circ$ and b) $240^\circ$ (see Figs.~\ref{chir} a) and b), respectively, for $XY$ TAF). Chiralities of such triangles are directed oppositely. In Heisenberg TAFs spins not necessarily lie in the plane of the lattice. In this case the chirality of a triangle can have any direction but the spin configuration with the opposite chirality also exists. By convention we assume that triangle a) has positive chirality and triangle b) has the negative one.

\begin{figure}
\centering
\includegraphics[scale=0.7]{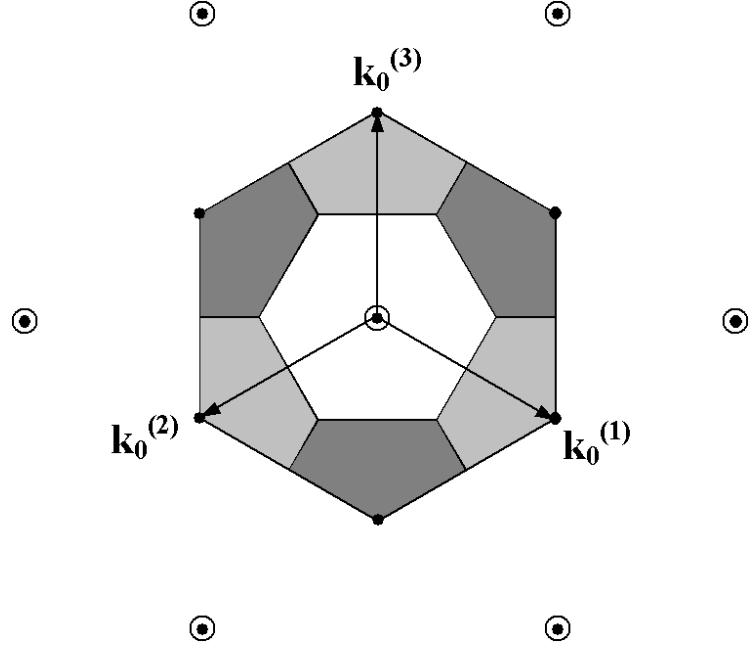}
\caption{
The reciprocal chemical lattice (RCL) (open circles) and the reciprocal magnetic lattice (RML) (black dots). The inner (white) hexagon is the first Brillouin zone (BZ) of RML. The outer hexagon defines the edge of the second BZ of RML and, simultaneously, the first BZ of RCL. Antiferromagnetic vectors ${\bf k}_0^{(1)}$, ${\bf k}_0^{(2)}$ and ${\bf k}_0^{(3)}$ are also presented. They are equivalent up to vectors of RCL. Also shown are two parts of the second BZ of RML which can be reduced to the first BZ by shifting on ${\bf k}_0$ (dark colored region) and $-{\bf k}_0$ (light colored region).
\label{zones}} 
\end{figure}

It is easy to show that there are two inequivalent $120^\circ$ spin structures shown in Fig.~\ref{chir} c) and d) which differ one from another by chiralities of each triangle. The spin order in configuration c) is determined by \cite{kawa2}
\begin{equation}
\label{s}
{\bf S}_{\bf R} = {\bf a} \cos ({\bf k}_0 {\bf R}) + {\bf b} \sin ({\bf k}_0{\bf R}),
\end{equation}
where ${\bf a}^2={\bf b}^2=S^2$, $({\bf a}{\bf b})=0$ and ${\bf k}_0$ is an antiferromagnetic vector which could be written in three forms (see Fig.~\ref{zones}):
\begin{equation}
\label{k0}
{\bf k}_0^{(1)} = 2\pi\left(0,\frac23,0\right), 
\qquad 
{\bf k}_0^{(2)} = 2\pi\left(\frac{1}{\sqrt{3}},-\frac13,0\right),
\qquad 
{\bf k}_0^{(3)} = 2\pi\left(-\frac{1}{\sqrt{3}},-\frac13,0\right),
\end{equation}
where the chemical lattice constant is taken equal to unity. As is seen from Fig.~\ref{zones}, they are equivalent up to the vectors of reciprocal chemical lattice. Spin configuration shown in Fig.~\ref{chir} d) is described by Eq.~(\ref{s}) with $-{\bf k}_0$ put instead of ${\bf k}_0$. Subsequently, configurations c) and d) are referred to as $(+{\bf k}_0)$ and $(-{\bf k}_0)$, respectively.

It is clear that $XY$ TAF has two inequivalent ground states which differ by chiralities of each elementary triangle. This degeneracy of the ground state of $XY$ TAF is called chiral degeneracy. As a result the symmetry of the order parameter is Z$_2\times$S$_1$ instead of S$_1$ for collinear $XY$ AFs, where Z$_2$ is the two-element group. In Heisenberg TAF the ground states $(+{\bf k}_0)$ and $(-{\bf k}_0)$ can be mutually transformed by global spin rotation. Nevertheless, the symmetry of the order parameter differs from that of non-frustrated AFs, as well. This is clear from the fact that the ground state in collinear AF is invariant under global spin rotation around magnetization of a sublattice whereas there is no such invariance in Heisenberg TAF. In Ref.~\cite{kawa} it is demonstrated that the symmetry of the order parameter in Heisenberg TAF is SO(3) instead of expected S$_2$.

Critical indexes of the chiral universality class calculated by Monte-Carlo simulations, $\epsilon$- and $1/n$- expansions are found to be different from those of Heisenberg magnets, $XY$ ferro- and $XY$ antiferromagnets. \cite{kawa} The largest deviation is in the index of specific heat: $\alpha=0.24\pm0.08$ ($n=3$) and $\alpha=0.34\pm0.06$ ($n=2$) for the chiral universality class and $\alpha\approx-0.12$ and $\alpha\approx-0.02$, respectively, for usual magnets. These findings of the theory have been confirmed in numerous experiments. \cite{kawa}

It should be noted an important role of polarized neutron scattering in investigation of the spin chirality. Nowadays, it gives the only opportunity to verify one of the most intriguing results by Kawamura that the spin chirality (\ref{c0}) is a new critical variable with fluctuations characterized by new critical indexes $\beta_c$ and $\gamma_c$. The former describes the temperature behavior of the average $\bf C$ and the latter characterizes its fluctuations near the transition temperature. Index $\beta_c$ can be investigated by elastic polarized neutron scattering. \cite{malbar,blume} Experimental observation of $\gamma_c$ is a more complicated task. As $\bf C$ is a composite two-spin operator, the corresponding fluctuations are described by four-spin susceptibility and their observation is hardly possible. At the same time, as is shown in Refs.~\cite{mal,mal2}, in presence of an axial vector interaction (e.g., Zeeman) the projection of $\bf C$ on it can be studied by polarized neutrons via the chiral term appearing in the cross section of inelastic scattering.

Experimentally, the chiral critical indexes in $XY$ TAFs have been determined recently for two members of the family of hexagonal AFs $ABX_3$ ($A$ = Rb, Cs; $B$ = Mn, Fe, Co, Ni, Cu; $X$ = Cl, Br, I), where magnetic atoms form triangular lattice: CsMnBr$_3$ ($S=5/2$) \cite{pl1} and CsNiCl$_3$ ($S=1$). \cite{pl2} The values of $\beta_c = 0.44(2)$ and $\gamma_c = 0.84(7)$ obtained for CsMnBr$_3$ and the sum $\beta_c+\gamma_c = 1.24(7)$ obtained for CsNiCl$_3$ are in a good agreement with the theory.

Thus, chiral fluctuations in TAFs near transition temperature have been extensively investigated recently and a certain success has been achieved. However chiral fluctuations in TAFs at temperatures much lower than the transition one have not been studied yet. One would expect that they have unusual properties in this case as well. 

In the present paper we consider chiral fluctuations in a quantum TAF CsCuCl$_3$ at $T \ll T_N$. The spin-wave spectrum of this compound has been studied, e.g., in Ref.~\cite{nikuni}. However, the chiral susceptibility has not been addressed yet. Such an analysis would be important for future investigation of CsCuCl$_3$ and other TAFs by polarized neutrons. 

CsCuCl$_3$ is a unique member of the family of hexagonal AFs $ABX_3$. Along $c$ direction the copper atoms form ferromagnetic chains whilst in $ab$ plane they interact antiferromagnetically and constitute a triangular lattice. The in-plane exchange is six times smaller than that along the chains. In contrast to CsMnBr$_3$ and CsNiCl$_3$ studied before, CsCuCl$_3$ is characterized by Dzyaloshinskii-Moriya (DM) interaction along the chains. This compound exhibits a phase transition at $T_N=10.7$ K to the ordered state wherein spins lie in $ab$ plane with $120^\circ$ structure. They form a helical structure along the chains with a pitch $5.1^\circ$ that is explained by a competition between ferromagnetic interaction within the chains and DM interaction. \cite{adachi} We will assume also that a small magnetic field ${\bf H} \| c$ is applied ($H\ll H_S$, where $H_S$ is a saturation field at which all spins become parallel).

In this paper expressions for the dynamical chirality (or chiral vector) are derived within the linear spin-wave approximation. We find that in contrast to the non-frustrated AFs it is finite at $D,H=0$. As is well known, \cite{mal,mal4} the chiral vector is nonzero only in the presence of an axial vector (e.g., magnetization of the sample, DM interaction, etc.) along which it is directed. We demonstrate that a ground state of $XY$ and Heisenberg TAFs is characterized by the axial vector. This is the consequence of existence of ground states $(+{\bf k}_0)$ and $(-{\bf k}_0)$. We show that these states are characterized by axial vectors $[{\bf a} \times {\bf b}]$ and $-[ {\bf a} \times {\bf b}]$, respectively. Hence, the chiral vector is perpendicular to the plane in which the spins lie and is directed oppositely for $(+{\bf k}_0)$ and $(-{\bf k}_0)$ configurations. Therefore, the dynamical chirality can be observed in a sample with a different population of domains. 

The paper is organized as follows. Main principles of experimental investigation of chiral fluctuations by polarized neutron are considered in Sec.~\ref{neutron}. We present there expressions for the cross section, its chiral part and the dynamical chirality. We discuss general transformations of the Hamiltonian describing CsCuCl$_3$ in Sec.~\ref{gr}. Green's functions properties and spin-wave spectrum of this compound are studied in Sec.~\ref{sw}. General properties of the chiral fluctuations in TAFs at $T\ll T_N$ are discussed in Sec.~\ref{genprop}. Expressions for chiral vector in CsCuCl$_3$ are derived in Sec.~\ref{chiralvec} within the linear spin-wave approximation. The range of validity of this approximation is also discussed there. Conclusions are made in Sec.~\ref{con}.

\section{Investigation of chiral fluctuations by polarized neutron}
\label{neutron}

We consider in this section in some detail the main principles of experimental investigation of chiral fluctuations by polarized neutron. For the neutron scattering amplitude we have the well-known expressions: \cite{low}
\begin{eqnarray}
F_{\bf Q} &=& N_{\bf Q}+{\bf M}_{\bf Q} \mbox{\boldmath $\sigma$},\label{f}\\
N_{\bf Q} &=& -\frac{1}{\sqrt { N}}\sum_n b_ne^{i{\bf Q}{\bf R}_n},\label{n}\\
{\bf M}_{\bf Q} &=& -\frac{1}{\sqrt { N}} \sum_m r F_m({\bf Q}) e^{i{\bf Q}{\bf R}_m}
[{\bf S}_m-(\hat Q{\bf S}_m)\hat Q], \label{m}
\end{eqnarray}
where $N$ is the total number of unit cells, $b_n$ are nuclear scattering lengths, $r=5.39\times 10^{-13}$ cm, $F_m({\bf Q})$ are magnetic form-factors and $\hat Q={\bf Q}/Q$, where ${\bf Q}$ is the momentum transfer. Four contributions to the inelastic scattering cross section can be represented using generalized retarded susceptibilities which have the form: \cite{mal4}
$
\langle A,B\rangle_\omega=\langle A,B\rangle '_\omega + i\langle
A,B\rangle ''_\omega,
$
where the first and the second terms are dispersive and absorptive parts, respectively. As a result for the cross section of inelastic scattering we have the following expression: \cite{mal}
\begin{equation}
\label{sigma}
\frac{d^2\sigma}{d\Omega d\omega} = 
\frac1\pi \frac{k_f}{k_i}
\frac{1}{1-\exp(-\omega/T)}
\left[
\langle N_{-\bf Q},N_{\bf Q}\rangle_\omega'' 
+
\langle {\bf M}_{-\bf Q}, {\bf M}_{\bf Q}\rangle_\omega''
+
i{\bf P}_0\langle [{\bf M}_{-\bf Q} \times {\bf M}_{\bf Q}]\rangle_\omega''
+
{\bf P}_0\langle N_{-\bf Q},{\bf M}_{\bf Q} + {\bf M}_{-\bf Q}, N_{\bf Q}\rangle_\omega''
\right],
\end{equation}
where ${\bf P}_0$ is the initial neutron polarization. The first two terms in Eq.~(\ref{sigma}) are nuclear and conventional magnetic parts of the cross section, respectively. The third and the fourth terms determine polarization-dependent part of the cross section that corresponds to chiral magnetic scattering and nuclear-magnetic interference, respectively. \cite{mal} As is seen from Eqs.~(\ref{m}) and (\ref{sigma}), one can eliminate contribution of the interference by directing polarization ${\bf P}_0$ along the impulse transfer $\bf Q$. In this case the chiral part of the scattering is obtained experimentally by subtracting results of the cross section measurements with polarization ${\bf P}_0$ and $-{\bf P}_0$. The chiral term in Eq.~(\ref{sigma}) is proportional to imaginary part of dynamical chirality ${\bf C}(\omega,{\bf Q})$ which is defined from the antisymmetric part of non-diagonal components of the generalized susceptibility: \cite{mal}
\begin{equation}
\label{sgf}
\chi_{\alpha\beta}(\omega,{\bf Q}) = \left\langle S_{-\bf Q}^\alpha, S_{\bf Q}^\beta\right\rangle_\omega
=
i\int_0^\infty dt e^{i\omega t}\left\langle\left[S_{-\bf Q}^\alpha(t), S_{\bf Q}^\beta(0)\right]\right\rangle,
\end{equation}
where $\alpha,\beta={x,y,z}$ and $\langle\dots\rangle$ denotes thermal average. Multiplying this expression by $i\epsilon_{\alpha\beta\gamma}$ we obtain the dynamical chirality:
\begin{equation}
\label{chirality}
{\bf C}(\omega,{\bf Q}) 
=
-\frac12\int_0^\infty dt e^{i\omega t}
\left\langle
\left[{\bf S}_{-\bf Q}(t)\times {\bf S}_{\bf Q}(0)\right] +
\left[{\bf S}_{\bf Q}(0)\times {\bf S}_{-\bf Q}(t)\right]
\right\rangle.
\end{equation}
As a result we have for the chiral part of the cross section from Eqs.~(\ref{m}) and (\ref{sigma}):
\begin{equation}
\label{sigmachir}
\frac{d^2\sigma_{ch}}{d\Omega d\omega} = 
\frac1\pi \frac{k_f}{k_i}
\frac{1}{1-\exp(-\omega/T)}
2r^2F^2_m ({\bf Q})({\bf P}_0\hat Q) (\hat Q {\rm Im}{\bf C}(\omega,{\bf Q})).
\end{equation}
Then, the main object of discussion of the present paper is the chiral vector ${\bf C}(\omega,{\bf Q})$.

For the elastic scattering cross section we have: \cite{mal}
\begin{eqnarray}
\label{sigmael}
\frac{d\sigma}{d\Omega} &=& 
\langle N_{-\bf Q}\rangle \langle N_{\bf Q}\rangle 
+
\langle {\bf M}_{-\bf Q}\rangle \langle {\bf M}_{\bf Q}\rangle
+
i{\bf P}_0[\langle {\bf M}_{-\bf Q} \rangle \times \langle {\bf M}_{\bf Q} \rangle]
+
{\bf P}_0(\langle N_{-\bf Q}\rangle \langle {\bf M}_{\bf Q}\rangle + \langle {\bf M}_{-\bf Q}\rangle \langle N_{\bf Q}\rangle).
\end{eqnarray}
One can express the chiral term in Eq.~(\ref{sigmael}) via the staggered chirality: \cite{lee}
\begin{equation}
\label{schir}
{\bf C}({\bf Q}) = 
i[{\bf S}_{\bf Q} \times {\bf S}_{-\bf Q}].
\end{equation}
We have from Eqs.~(\ref{s}) and (\ref{sigmael}) that the chiral term in the elastic scattering cross section is proportional to ${\bf C}^\perp({\bf Q}){\bf P}_0$, where
\begin{equation}
\label{cs}
{\bf C}^\perp({\bf Q}) = \frac12 \hat Q \left\langle \left([{\bf a} \times {\bf b}]\cdot\hat Q\right) \right\rangle \sum_{\mbox{\boldmath $\tau$}}
[\varDelta({\bf Q} - {\bf k}_0 + \mbox{\boldmath $\tau$})
-
\varDelta({\bf Q} + {\bf k}_0 + \mbox{\boldmath $\tau$})],
\end{equation}
where $\mbox{\boldmath $\tau$}$ is the reciprocal lattice vector, $\varDelta({\bf 0})=N$, $\varDelta({\bf Q}\ne0)=0$ and $\hat Q = {\bf Q}/Q$. Notice that the dynamical chirality given by Eq.~(\ref{chirality}) is a natural generalization of the staggered one given by Eq.~(\ref{schir}). It is seen from Eq.~(\ref{cs}) that ${\bf C}^\perp({\bf Q})$ is expressed via the average static chirality which can be investigated near Bragg reflections. Since Eq.~(\ref{cs}) is an odd function of ${\bf k}_0$, the polarization-dependent part of the cross section has opposite signs for states described by ${\bf k}_0$ and $-{\bf k}_0$. However, if the energy does not depend on the sign of antiferromagnetic vector (as in the case of centrosymmetric crystals CsMnBr$_3$, CsNiCl$_3$ and helimagnetic holmium) the sample splits onto domains with ${\bf k}_0$ and $-{\bf k}_0$. Therefore, the chiral scattering will be proportional to the difference of their population and can be observed only if this difference is nonzero. For CsMnBr$_3$ it is the case. \cite{pl1} For helimagnetic holmium the difference in domain population has been achieved by cooling of the twisted crystal below the transition temperature. \cite{pl3} 

\section{Model Hamiltonian}
\label{gr}

The model Hamiltonian of CsCuCl$_3$ has the form: \cite{nikuni}
\begin{equation}
\label{ham}
{\cal H} = -2J_0\sum_{in}
\left(
{\bf S}_{in}{\bf S}_{in+1} 
+ \eta (S_{in}^x S_{in+1}^x + S_{in}^y S_{in+1}^y)
\right)
+ 2J_1\sum_{\langle ij\rangle n}{\bf S}_{in}{\bf S}_{jn}
-\sum_{in}{\bf D}\cdot [{\bf S}_{in}\times {\bf S}_{in+1}]
-g\mu_BH\sum_{in}S_{in}^z,
\end{equation}
where ${\bf S}_{in}$ is operator of a spin at the $i$-th site in the $n$-th $ab$ plane, $z$ axis is taken to be parallel to $c$ axis, $J_0>0$ and $J_1>0$ are values of exchange along $c$ axis and within $ab$ plane, respectively, $0<\eta\ll1$ is a weak in-plane anisotropy, $\langle ij\rangle$ denote nearest-neighbor spins in $ab$ plane. The third term in (\ref{ham}) describes DM interaction with vector $\bf D$ being parallel to $c$ axis. The fourth term is the Zeeman energy. Coefficients in (\ref{ham}) have been estimated previously (see Refs.~\cite{tanaka,nikuni} and references therein): $J_0\approx28$ K, $J_1\approx4.9$ K, $D\approx5$ K and $\eta\approx0.008$.

One can simplify the expression for $\cal H$ by applying a standard transformation of spin operators: \cite{nikuni,aristov}
\begin{equation}
\label{tr}
\left\{
\begin{array}{l}
S_{in}^x = S_{in}^{x\prime}\cos nq - S_{in}^{y\prime}\sin nq\\
S_{in}^y = S_{in}^{x\prime}\sin nq + S_{in}^{y\prime}\cos nq
\end{array}
\right.
\end{equation}
which implies a rotation of $xy$ plane by a pitch $q$ along $z$ axis. The value of $q$ is determined so as to eliminate the antisymmetric part of the Hamiltonian $[{\bf S}_{in}\times {\bf S}_{in+1}]$: 
\begin{equation}
\label{q}
\tan q = \frac{D}{2J_0(1+\eta)}. 
\end{equation}
As a result the Hamiltonian has the form:
\begin{equation}
\label{hamtr}
{\cal H} = -\sum_{in}\left(2\tilde J_0 (S_{in}^{x\prime} S_{in+1}^{x\prime} + S_{in}^{y\prime} S_{in+1}^{y\prime})
+ 2J_0 S_{in}^{z\prime} S_{in+1}^{z\prime}
\right)
+ 2J_1\sum_{\langle ij\rangle n}{\bf S}_{in}'{\bf S}_{jn}'
-g\mu_BH\sum_{in}S_{in}^{z\prime},
\end{equation}
where 
\begin{equation}
\label{j}
\tilde J_0 = J_0\sqrt{(1+\eta)^2+\left(\frac{D}{2J_0}\right)^2}\approx1.012J_0.
\end{equation}
This way DM interaction has been reduced to a contribution to in-plane anisotropy. As is shown in Ref.~\cite{nikuni}, at $H=0$ this anisotropy favors the coplanar ground state configuration with three sublattices and $120^\circ$ spin structure. Spins are canted by magnetic field forming at $H\ll H_S$ an umbrella-like configuration. In the following derivations we assume this spin structure of the ground state and neglect the small difference between $\tilde J_0$ and $J_0$. 

For consideration of spin waves it is convenient to use the following representations for the spin operators:
\begin{eqnarray}
\label{rep1}
&&\left\{
\begin{array}{l}
S_{in}^{x\prime} = -h\cos({\bf k}_0{\bf R}_{in})S_{in}^{x\prime\prime} - \sin({\bf k}_0{\bf R}_{in})S_{in}^{y\prime\prime} + \sqrt{1-h^2}\cos({\bf k}_0{\bf R}_{in})S_{in}^{z\prime\prime},\\
S_{in}^{y\prime} = -h\sin({\bf k}_0{\bf R}_{in})S_{in}^{x\prime\prime} + \cos({\bf k}_0{\bf R}_{in})S_{in}^{y\prime\prime} + \sqrt{1-h^2}\sin({\bf k}_0{\bf R}_{in})S_{in}^{z\prime\prime}, \\
S_{in}^{z\prime} = -\sqrt{1-h^2}S_{in}^{x\prime\prime} - hS_{in}^{z\prime\prime},
\end{array}
\right.
\\
\label{rep2}
&&\left\{
\begin{array}{l}
S_{in}^{x\prime\prime} = \sqrt{\frac S2}(a_{in}+a_{in}^\dagger - \frac{a^\dagger_{in}a_{in}^2}{2S}),\\
S_{in}^{y\prime\prime} = -i\sqrt{\frac S2}(a_{in}-a_{in}^\dagger - \frac{a^\dagger_{in}a_{in}^2}{2S}),\\ 
S_{in}^{z\prime\prime} = S-a^\dagger_{in}a_{in},
\end{array}
\right.
\end{eqnarray}
where $h \ge 0$ is a sinus of the canting angle ($h=0$ at $H=0$), derived below, and ${\bf k}_0$ is the antiferromagnetic vector. It should be stressed that the spin transformation (\ref{rep1}) with antiferromagnetic vectors ${\bf k}_0$ given by Eq.~(\ref{k0}) describes configuration $(+{\bf k}_0)$ (Fig.~\ref{chir} c)). One has to use $-{\bf k}_0$ instead of ${\bf k}_0$ in (\ref{rep1}) to describe configuration $(-{\bf k}_0)$ (Fig.~\ref{chir} d)). This difference has no effect on results of spin-wave spectrum calculations but, as is shown below, it is of great importance for chiral fluctuations. Unless otherwise specified, all formulas presented below are for $(+{\bf k}_0)$ configuration.

Note that representations (\ref{rep1}) and (\ref{rep2}) contain only one type of Bose operators. This makes intermediate calculations and results more compact. Such an approach has been previously applied for other problems \cite{pet,afminh} and proved to be more convenient than the traditional one dealing with several types of Bose operators ascribed to different sublattices.

Substituting Eq.~(\ref{rep1}) into Eq.~(\ref{hamtr}) and taking the Fourier transform of spin operators: 
\begin{equation}
\label{four}
{\bf S}_{in}'' = \sqrt{\frac1N}\sum_{\bf k} {\bf S}_{\bf k}'' e^{i{\bf k R}_{in}},
\end{equation}
where $N$ is the number of spins in the lattice and the sum is over the chemical Brillouin zone (BZ) (see Fig.~\ref{zones}), we have for the Hamiltonian:
\begin{equation}
\label{hamtr2}
{\cal H} = -2J_0\sum_{\bf k} \cos k_z{\bf S}_{\bf k}''{\bf S}_{-\bf k}''
+ 6J_1\sum_{\bf k} \nu_{\bf k}{\bf S}_{\bf k}''{\bf S}_{-\bf k}'' 
- g\mu_BH\sqrt{N}S_{\bf 0}^{z\prime\prime},
\end{equation}
where 
\begin{equation}
\label{nu}
\nu_{\bf k} = \frac13
\left(
\exp\{ik_y\} + \exp\left\{i\left(\frac{\sqrt{3}}{2}k_x - \frac12 k_y\right)\right\} + \exp\left\{i\left(-\frac{\sqrt{3}}{2}k_x - \frac12 k_y\right)\right\}
\right).
\end{equation}
Let us substitute Eq.~(\ref{rep2}) into Eq.~(\ref{hamtr2}). The classical ground state energy is given by the term without Bose operators:
\begin{equation}
\label{ground}
E_0 = NS(g\mu_BhH - 2SJ_0 - 3(1-3h^2)SJ_1).
\end{equation}
Minimization of Eq.~(\ref{ground}) with respect to $h$ yields $h=-g\mu_BH/H_S$ ($H_S=18SJ_1$ is the saturation field) in accordance with Ref.~\cite{nikuni}. In this case the terms linear in Bose operators cancel each other and the Hamiltonian has the form ${\cal H} = E_0 + \sum_{i=2}^{6} {\cal H}_i$, where ${\cal H}_i$ describes terms with products of $i$ operators $a$ and $a^\dagger$. Below we derive the spin chirality within the linear spin-wave approximation, i.e., we restrict ourself to bilinear part of the Hamiltonian which has a simple form:
\begin{eqnarray}
\label{h2}
&&{\cal H}_2 = \sum_{\bf k} \left[
E_{\bf k} a^\dagger_{\bf k}a_{\bf k} 
+
\frac{B_{\bf k}}{2}(a_{\bf k}a_{-\bf k} + a_{\bf k}^\dagger a_{-\bf k}^\dagger)\right],\nonumber\\
&& E_{\bf k} = 4SJ_0(1-\cos k_z) + 6SJ_1 + 3SJ_1(1-3h^2)\lambda_{\bf k} -6\sqrt{3}SJ_1h\xi_{\bf k}, \\
&& B_{\bf k} = 9SJ_1(1-h^2)\lambda_{\bf k}\nonumber
\end{eqnarray}
where $\lambda_{\bf k} = {\rm Re}\, \nu_{\bf k}$ and  $\xi_{\bf k} = {\rm Im}\, \nu_{\bf k}$. The range of validity of this approximation is discussed at the end of Sec.~\ref{chiralvec}.

\section{Green's functions and spin-wave spectrum}
\label{sw}

It is convenient to introduce two Green's functions: $G(\omega,{\bf k}) = -\langle a_{\bf k},a_{\bf k}^\dagger\rangle_\omega$ and $F(\omega,{\bf k}) = -\langle a_{\bf k},a_{-\bf k} \rangle_\omega$ (definition of $\langle \dots \rangle_\omega$ is given in Eq.~(\ref{sgf})). Spin-wave spectrum can be obtained from analysis of their denominator. There are two other Green's functions connected to the ones above: $\bar G(\omega,{\bf k}) = -\langle a_{-\bf k}^\dagger,a_{-\bf k}\rangle_\omega=G^*(-\omega,-{\bf k})$ and $F^\dagger(\omega,{\bf k}) = -\langle a^\dagger_{-\bf k},a^\dagger_{\bf k} \rangle_\omega=F^*(-\omega,-{\bf k})$. As a result we have two sets of Dyson's equations. One of them has the following form:
\begin{equation}
\label{eqfunc}
\begin{array}{l}
G(\omega,{\bf k}) = G^{(0)}(\omega,{\bf k}) + G^{(0)}(\omega,{\bf k}) B_{\bf k} F^\dagger(\omega,{\bf k}),\\
F^\dagger(\omega,{\bf k}) = \bar G^{(0)}(\omega,{\bf k}) B_{\bf k} G(\omega,{\bf k}),
\end{array}
\end{equation}
where $G^{(0)}(\omega,{\bf k}) = (\omega - E_{\bf k}+i\delta)^{-1}$ is the bare Green's function. Solving Eqs.~(\ref{eqfunc}) we have:
\begin{equation}
\label{grf}
G(\omega,{\bf k}) = \frac{\omega + E_{-\bf k}}{\Delta(\omega,{\bf k})}, 
\qquad 
\bar G(\omega,{\bf k}) = \frac{-\omega + E_{\bf k}}{\Delta(\omega,{\bf k})}, 
\qquad 
F^\dagger(\omega,{\bf k}) = F(\omega,{\bf k}) = -\frac{B_{\bf k}}{\Delta(\omega,{\bf k})},
\end{equation}
where
\begin{equation}
\label{delta}
\Delta(\omega,{\bf k}) = (\omega+i\delta)^2 - (\omega+i\delta)(E_{\bf k} - E_{-\bf k}) - E_{\bf k}E_{-\bf k} + B_{\bf k}^2 
= (\omega - \epsilon_{\bf k} + i\delta)(\omega + \epsilon_{-\bf k} + i\delta).
\end{equation}
Here
\begin{equation}
\label{spec}
\epsilon_{\bf k} = \sqrt{\bigl[4SJ_0(1-\cos k_z) + 6SJ_1(1-\lambda_{\bf k})\bigr]\bigl[4SJ_0(1-\cos k_z) + 6SJ_1\{1+(2-3h^2)\lambda_{\bf k}\}\bigr]} - 6\sqrt{3}SJ_1h\xi_{\bf k}
\end{equation}
determines the spin-wave spectrum which is known to have three branches. \cite{nikuni} Eq.~(\ref{spec}) describes all three branches at once because vector $\bf k$ in our consideration lies within the first BZ of the reciprocal chemical lattice which includes the first and the second BZ of the reciprocal magnetic lattice (RML) (see Fig.~\ref{zones}). Consequently, the first branch of the spectrum is given by Eq.~(\ref{spec}) with {\bf k} within the first BZ of RML. The second and the third branches are described by  Eq.~(\ref{spec}) with {\bf k} being within, respectively, dark colored and light colored areas of the second BZ of RML shown in Fig.~\ref{zones}. One can represent expressions for these branches in a traditional form writing
\begin{equation}
\label{branches}
\epsilon_{\bf k}^{(1)} = \epsilon_{\bf k},
\qquad
\epsilon_{\bf k}^{(2)} = \epsilon_{{\bf k} + {\bf k}_0},
\qquad
\epsilon_{\bf k}^{(3)} = \epsilon_{{\bf k} - {\bf k}_0},
\end{equation}
where $\epsilon_{\bf k}$ is given by Eq.~(\ref{spec}) and vector {\bf k} lies within the first BZ of the RML. It can be easily verified using Eqs.~(\ref{k0}), (\ref{nu}), (\ref{h2}) and (\ref{spec}) that these expressions are in agreement with those of Ref.~\cite{nikuni}. 

In the vicinity of points ${\bf k}=0$, ${\bf k}={\bf k}_0$ and ${\bf k}=-{\bf k}_0$ the spectrum has the form:
\begin{equation}
\label{ssk}
\epsilon_{\bf k}^2=
\left\{
\begin{array}{ll}
\left(c_\|^{(0)} k_z\right)^2 + \left(c_\perp^{(0)} k_\perp\right)^2 & \mbox{\quad if } k\ll k_0, \\
\left(c_\|^{(k_0)} k_z\right)^2 + \left(c_\perp^{(k_0)}\right)^2 \left({\bf k}_\perp-{\bf k}_0\right)^2 + \Delta^2   & \mbox{\quad if } |{\bf k}-{\bf k}_0|\ll k_0,\\
\left(c_\|^{(k_0)} k_z\right)^2 + \left(c_\perp^{(k_0)}\right)^2 \left({\bf k}_\perp+{\bf k}_0\right)^2 & \mbox{\quad if } |{\bf k}+{\bf k}_0|\ll k_0,
\end{array}
\right.
\end{equation}
where $c_\|^{(0)}=\sqrt2 c_\|^{(k_0)}=6S\sqrt{J_0J_1}$,  $c_\perp^{(0)}=\sqrt2 c_\perp^{(k_0)}=3\sqrt{3}SJ_1$, ${\bf k}_\perp=(k_x,k_y,0)$ and $\Delta = 18SJ_1h$. We see that the spectrum is linear near these points with different spin-wave velocities at $k\ll k_0$ and $|{\bf k}\pm{\bf k}_0|\ll k_0$. Magnetic field leads to a gap $\Delta$ at ${\bf k}={\bf k}_0$. We neglect here a small gap related to DM interaction. In our description it is proportional to $\tilde J_0 - J_0$ (see Eq.~(\ref{j})).

\section{Chiral fluctuations}
\label{chf}
We discuss in this section general properties of the chiral fluctuations in TAFs and derive the dynamical chirality for CsCuCl$_3$. 

\subsection{General properties}
\label{genprop}

As is well known, \cite{mal,mal4} the chiral vector is nonzero only in the presence of an axial vector (e.g.\ magnetization of the sample, DM interaction etc.) and is directed along it. To illustrate this point we consider two examples: square antiferromagnet and ferromagnet.

Chiral vector is zero in the first case. To show that $C_y=0$ ($z$ axis is along magnetization of sublattices) we make a rotation of the coordinate space by an angle $\pi$ along $x$ direction. The Hamiltonian does not change under this transformation and the ground state spin configuration is described by the same antiferromagnetic vector. Hence from Eq.~(\ref{chirality}) we have $C_y(\omega,{\bf Q})=-C_y(\omega,{\bf Q})=0$. One can show in the same way that $C_z=C_x=0$.

In the case of a ferromagnet the chiral vector is directed along magnetization of the sample (one-domain case is considered for simplicity). It becomes clear if we direct $z$ axis along magnetization $\bf m$ and make the rotation of the coordinate space by $\pi$ along $x$ direction. The Hamiltonian does not change and the ground state wave function is transformed to that of state with magnetization of the opposite direction. Thus from (\ref{chirality}) we have $C_z^{({\bf m})}(\omega,{\bf Q})=-C_z^{(-{\bf m})}(\omega,{\bf Q})$. Making the rotation along $z$ direction it is easy to show that $C_y=C_x=0$. Then, the chiral vector is directed along ${\bf m}$.

We demonstrate now that there is a different situation in $XY$ and Heisenberg TAFs. Let us assume that the ground state has configuration $(+{\bf k}_0)$ and $z$ axis is directed perpendicular to the plane in which the spins lie. It is easy to see that after the rotation of the coordinate space by $\pi$ along $x$ direction the Hamiltonian does not change and the ground state spin configuration becomes $(-{\bf k}_0)$. As a result we have $C_z^{(+{\bf k}_0)}(\omega,{\bf Q})=-C_z^{(-{\bf k}_0)}(\omega,{\bf Q})$. The rotation along $z$ direction gives $C_y=C_x=0$. It means that chiral vector in configuration $(+{\bf k}_0)$ has the same value as and is directed oppositely to that in configuration $(-{\bf k}_0)$. But it certainly does not mean that it is zero. Results of straightforward calculations for $XY$ TAF presented below show that it is in fact nonzero. 

Thus we see that existence of inequivalent ground states described by Eq.~(\ref{s}) with the same $\bf a$ and $\bf b$ and different antiferromagnetic vectors ${\bf k}_0$ and $-{\bf k}_0$ is the origin of this unusual properties of chiral fluctuations. As a result direction perpendicular to the plane in which spins lie is physically selected. It is clear that the staggered chirality at ${\bf Q}={\bf k}_0$ can play the role of the axial vector characterizing the system. From Eqs.~(\ref{s}) and (\ref{schir}), ${\bf C}({\bf k}_0) = \frac12 [ {\bf a} \times {\bf b}]$ for $(+{\bf k}_0)$ and ${\bf C}({\bf k}_0) = -\frac12 [ {\bf a} \times {\bf b}]$ for $(-{\bf k}_0)$ configurations.

It should be noted that in real substances there are domains with both spin configurations $(+{\bf k}_0)$ and $(-{\bf k}_0)$, so that we have to introduce concentrations $n_{(+{\bf k}_0)}$ and $n_{(-{\bf k}_0)}$ of these domains in the sample, with $n_{(+{\bf k}_0)} + n_{(-{\bf k}_0)}=1$. If $D,H=0$ their contribution to the chiral vector has opposite signs and, similar to staggered chirality (\ref{schir}), the dynamical one would be nonzero only when $n_{(+{\bf k}_0)} \ne n_{(-{\bf k}_0)}$.

It can be shown using symmetry with respect to time inversion and definition (\ref{chirality}) that the chiral vector in general has the following properties: \cite{mal,mal4}
\begin{eqnarray}
\label{sym1}
{\bf C}(\omega,{\bf Q},{\bf H}) &=& -{\bf C}(\omega,-{\bf Q},-{\bf H}),\\
\label{sym2}
{\rm Im}{\bf C}(\omega,{\bf Q},{\bf H}) &=& {\rm Im}{\bf C}(-\omega,-{\bf Q},{\bf H}) 
= -{\rm Im}{\bf C}(-\omega,{\bf Q},-{\bf H}).
\end{eqnarray}

\subsection{Chiral vector in CsCuCl$_3$}
\label{chiralvec}

We proceed with calculation of the dynamical chirality in CsCuCl$_3$. As is shown in Sec.~\ref{neutron}, it is expressed via non-diagonal components of spin Green's function (\ref{sgf}). In CsCuCl$_3$ there is only one non-diagonal component, $\chi_{xy}=-\chi_{yx}$, and the chiral vector (\ref{chirality}) is directed along $z$ axis:
\begin{equation}
\label{c}
C_z(\omega,{\bf Q}) = i\chi_{xy}(\omega,{\bf Q}).
\end{equation}
To calculate it within the linear spin-wave approximation one has to use Eqs.~(\ref{sgf}) and (\ref{c}), perform sequentially transformations of spin operators (\ref{tr}), (\ref{rep1}), (\ref{rep2}) and use expressions (\ref{grf}) for Green's functions. 

Let us consider firstly general expressions for the chiral vector. It is easy to show using Eq.~(\ref{four}) that after transformation (\ref{tr}) the spin operators have the form: 
\begin{eqnarray}
S_{\bf Q}^x &=& \frac12 (S_{{\bf Q}_-}^{x\prime} + S_{{\bf Q}_+}^{x\prime})
+\frac i2 (S_{{\bf Q}_-}^{y\prime} - S_{{\bf Q}_+}^{y\prime}),\nonumber\\
S_{\bf Q}^y &=& -\frac i2 (S_{{\bf Q}_-}^{x\prime} - S_{{\bf Q}_+}^{x\prime})
+\frac 12 (S_{{\bf Q}_-}^{y\prime} + S_{{\bf Q}_+}^{y\prime}),
\end{eqnarray}
where 
\begin{equation}
\label{qpm}
{\bf Q}_+ = {\bf Q} + q\hat c,
\qquad
{\bf Q}_- = {\bf Q} - q\hat c.
\end{equation}
Here $q$ is given by Eq.~(\ref{q}) and $\hat c$ is a unite vector directed along $c$ axis. After performing transformation (\ref{tr}), for the chiral vector we have:
\begin{equation}
\label{c1}
C_z(\omega,{\bf Q}) = 
\frac12\left[
\chi_{xx}'(\omega,{\bf Q_-})
- \chi_{xx}'(\omega,{\bf Q_+})
\right]
+
\frac i2\left[\chi_{xy}'(\omega,{\bf Q_-})
+ \chi_{xy}'(\omega,{\bf Q_+})
\right],
\end{equation}
where $\chi_{\alpha\beta}'(\omega,{\bf Q})=\left\langle S_{-{\bf Q}}^{\alpha\prime}, S_{{\bf Q}}^{\beta\prime}\right\rangle_\omega$. Deriving Eq.~(\ref{c1}) we also used $\chi_{xx}'(\omega,{\bf Q})=\chi_{yy}'(\omega,{\bf Q})$ and $\chi_{xy}'(\omega,{\bf Q})=-\chi_{yx}'(\omega,{\bf Q})$.
Note that the first term here is nonzero only when $q\ne0$ (i.e.\ $D\ne0$). In contrast to non-frustrated systems the second term in (\ref{c1}) and the chiral vector itself is nonzero at $D,H=0$. After transformation (\ref{rep1}) for $\chi_{xx}'(\omega,{\bf Q})$ and $\chi_{xy}'(\omega,{\bf Q})$ appearing in (\ref{c1}) we obtain:
\begin{eqnarray}
\label{xx}
\chi_{xx}'(\omega,{\bf Q})
&=&
\frac 14 
\left[  
\chi_{yy}''(\omega,{\bf Q} + {\bf k}_0)
+
\chi_{yy}''(\omega,{\bf Q} - {\bf k}_0)
+
h^2\left\{
\chi_{xx}''(\omega,{\bf Q} + {\bf k}_0)
+
\chi_{xx}''(\omega,{\bf Q} - {\bf k}_0)
\right\}
\right]\nonumber\\
&&{}+
\frac {ih}{2}
\left[ 
\chi_{xy}''(\omega,{\bf Q} + {\bf k}_0)
-
\chi_{xy}''(\omega,{\bf Q} - {\bf k}_0)
\right]\\
\label{xy}
\chi_{xy}'(\omega,{\bf Q})
&=&
\frac i4 
\left[  
\chi_{yy}''(\omega,{\bf Q} + {\bf k}_0)
-
\chi_{yy}''(\omega,{\bf Q} - {\bf k}_0)
+
h^2\left\{
\chi_{xx}''(\omega,{\bf Q} + {\bf k}_0)
-
\chi_{xx}''(\omega,{\bf Q} - {\bf k}_0)
\right\}
\right]\nonumber\\
&&{}-
\frac h2
\left[ 
\chi_{xy}''(\omega,{\bf Q} + {\bf k}_0)
+
\chi_{xy}''(\omega,{\bf Q} - {\bf k}_0)
\right],
\end{eqnarray}
where $\chi_{\alpha\beta}''(\omega,{\bf Q}) = \left\langle S_{-\bf Q}^{\alpha\prime\prime}, S_{\bf Q}^{\beta\prime\prime}\right\rangle_\omega$. To derive an explicit expression for chiral vector we have to make transformation (\ref{rep2}) and use Eq.~(\ref{grf}) for Green's functions. To calculate the corresponding expressions for configuration $(-{\bf k}_0)$ one has to replace ${\bf k}_0$ by $-{\bf k}_0$ in (\ref{xx}) and (\ref{xy}) and note that bilinear part of the Hamiltonian for $(-{\bf k}_0)$ has the form (\ref{h2}) with a formal replacement of $h$ by $-h$.

As a result of straightforward calculations, using $n_{(+{\bf k}_0)}=1/2+(n_{(+{\bf k}_0)}-n_{(-{\bf k}_0)})/2$ and  $n_{(-{\bf k}_0)}=1/2-(n_{(+{\bf k}_0)}-n_{(-{\bf k}_0)})/2$ and noting that terms proportional to $n_{(+{\bf k}_0)}-n_{(-{\bf k}_0)}$ cancel each other, we have for the chiral vector at $Q\ll k_0$:
\begin{equation}
\label{cz0}
C_z(\omega,{\bf Q}) =
-\frac{1}{72J_1}
\left[
\frac{\epsilon_{{\bf Q}_-}^2 + 36SJ_1\omega h}
{\Delta(\omega,{\bf Q}_- + {\bf k}_0)}
- \frac{\epsilon_{{\bf Q}_+}^2 - 36SJ_1\omega h}
{\Delta(\omega,{\bf Q}_+ - {\bf k}_0)}
\right].
\end{equation}
As is seen from Eq.~(\ref{cz0}), the chiral vector is zero at $D,H=0$ and $Q\ll k_0$.

For $|{\bf Q} - {\bf k}_0|\ll k_0$ one has:
\begin{eqnarray}
\label{czk0}
C_z(\omega,{\bf Q}) &=& 
-\frac S8
\left[
\frac{18SJ_1}{\Delta(\omega,{\bf Q}_- - {\bf k}_0)}
-
\frac{18SJ_1}{\Delta(\omega,{\bf Q}_+ - {\bf k}_0)}
+
\frac{\epsilon_{{\bf Q}_-}^2 + 18SJ_1\omega h}{9SJ_1\Delta(\omega,{\bf Q}_- + {\bf k}_0)}
-
\frac{\epsilon_{{\bf Q}_+}^2 - 18SJ_1\omega h}{9SJ_1\Delta(\omega,{\bf Q}_+)}
\right]
\nonumber\\
&&{}
-
\frac S8
[n_{(+{\bf k}_0)}-n_{(-{\bf k}_0)}]
\left[
\frac{18SJ_1}{\Delta(\omega,{\bf Q}_- - {\bf k}_0)}
+
\frac{18SJ_1}{\Delta(\omega,{\bf Q}_+ - {\bf k}_0)}
-
\frac{\epsilon_{{\bf Q}_-}^2 + 18SJ_1\omega h}{9SJ_1\Delta(\omega,{\bf Q}_- + {\bf k}_0)}
-
\frac{\epsilon_{{\bf Q}_+}^2 - 18SJ_1\omega h}{9SJ_1\Delta(\omega,{\bf Q}_+)}
\right].
\nonumber\\
\end{eqnarray}
The first term in Eq.~(\ref{czk0}) vanishes at $D,H=0$. The second term is proportional to $n_{(+{\bf k}_0)}-n_{(-{\bf k}_0)}$ and gives a finite contribution to chiral vector even at $D,H=0$ provided that $n_{(+{\bf k}_0)}\ne n_{(-{\bf k}_0)}$. In accordance with the results above, at $D,H=0$ the domains of the type $(+{\bf k}_0)$ and those of the type $(-{\bf k}_0)$ give contributions with opposite signs.

Finally, for $|{\bf Q} + {\bf k}_0|\ll k_0$ the chiral vector becomes:
\begin{eqnarray}
\label{cz-k0}
C_z(\omega,{\bf Q}) &=& 
-\frac S8
\left[
\frac{18SJ_1}{\Delta(\omega,{\bf Q}_- + {\bf k}_0)}
-
\frac{18SJ_1}{\Delta(\omega,{\bf Q}_+ + {\bf k}_0)}
+
\frac{\epsilon_{-{\bf Q}_-}^2 + 18SJ_1\omega h}{9SJ_1\Delta(\omega,{\bf Q}_-)}
-
\frac{\epsilon_{-{\bf Q}_+}^2 - 18SJ_1\omega h}{9SJ_1\Delta(\omega,{\bf Q}_+ - {\bf k}_0)}
\right]
\nonumber\\
&&{}
+
\frac S8
[n_{(+{\bf k}_0)}-n_{(-{\bf k}_0)}]
\left[
\frac{18SJ_1}{\Delta(\omega,{\bf Q}_- + {\bf k}_0)}
+
\frac{18SJ_1}{\Delta(\omega,{\bf Q}_+ + {\bf k}_0)}
-
\frac{\epsilon_{-{\bf Q}_-}^2 + 18SJ_1\omega h}{9SJ_1\Delta(\omega,{\bf Q}_-)}
-
\frac{\epsilon_{-{\bf Q}_+}^2 - 18SJ_1\omega h}{9SJ_1\Delta(\omega,{\bf Q}_+ - {\bf k}_0)}
\right].
\nonumber\\
\end{eqnarray}
This expression has the same structure as that for $|{\bf Q} - {\bf k}_0|\ll k_0$ discussed above.

As a result we see that even at $D,H=0$ the chiral vector remains finite in the vicinity of points ${\bf k}_0$ and $-{\bf k}_0$, as long as the population of domains of the type $(+{\bf k}_0)$ and $(-{\bf k}_0)$ are different.

Using expressions (\ref{cz0})--(\ref{cz-k0}) and
\begin{equation}
\label{im}
{\rm Im}\frac{1}{\Delta(\omega,{\bf k})} = 
\frac{\pi}{\epsilon_{\bf k} +\epsilon_{-\bf k}}
[
\delta(\omega+\epsilon_{-\bf k}) 
-
\delta(\omega-\epsilon_{\bf k}) 
]
\end{equation}
one can verify that relations (\ref{sym1}) and (\ref{sym2}) hold in our case. In particular, the terms in expressions for $\bf C$ and $\rm Im \bf C$ with different $h$-parity have different $\omega$- and $\bf Q$- parity. Indeed, according to (\ref{sym1}) terms in $\bf C$ which are odd functions of $h$ are even functions of $\bf Q$ and vice versa: even-$h$ terms are odd functions of $\bf Q$. Similarly, we determine from (\ref{sym2}) that even-$h$ and odd-$h$ terms in $\rm Im \bf C$ are odd and even functions of $\omega$, respectively.

Using Eqs.~(\ref{sigmachir}), (\ref{cz0})--(\ref{im}) it is easy to derive the chiral part of the cross section. Because of cumbersomeness of the resulting expressions we do not present them here. The only thing we note is that according to Eqs.~(\ref{sym2}) and (\ref{sigmachir}) at $\omega\gg T$ even-$h$ and odd-$h$ terms in the cross section are odd and even functions of $\omega$, respectively.

Let us consider now the range of validity of final expressions for the chiral vector (\ref{cz0})--(\ref{cz-k0}). To derive these expressions we used the linear spin-wave approximation restricting the initial Hamiltonian (\ref{ham}) after transformations  (\ref{tr}), (\ref{rep1}) and (\ref{rep2}) to its bilinear part in operators of creation and annihilation (\ref{h2}). This approximation is widely used because it is believed that $1/S$-corrections stemming from high-order terms in Hamiltonian are small. It is the case, e.g., for square and cubic Heisenberg AFs. \cite{kop,cast,canali,igar,harris,malold} At the same time the spin-wave interaction is important in some cases. For example, in square and cubic AFs in the magnetic field the first $1/S$-correction to the chiral vector become large at small $T$, $\omega$ and $Q$. \cite{afminh} We carried out the corresponding analysis for TAFs. We found that the first $1/S$-corrections to the spin-wave spectrum and to the chiral vector are small. Therefore, one can restrict ourselves by the results of linear spin-wave approximation (\ref{cz0})--(\ref{cz-k0}) in the range of validity of representation (\ref{rep2}), i.e., at $T\ll T_N$. The detailed discussion of this question is out of the scope of this paper.

\section{Conclusion}
\label{con}

\begin{figure}
\centering
\includegraphics[scale=0.7]{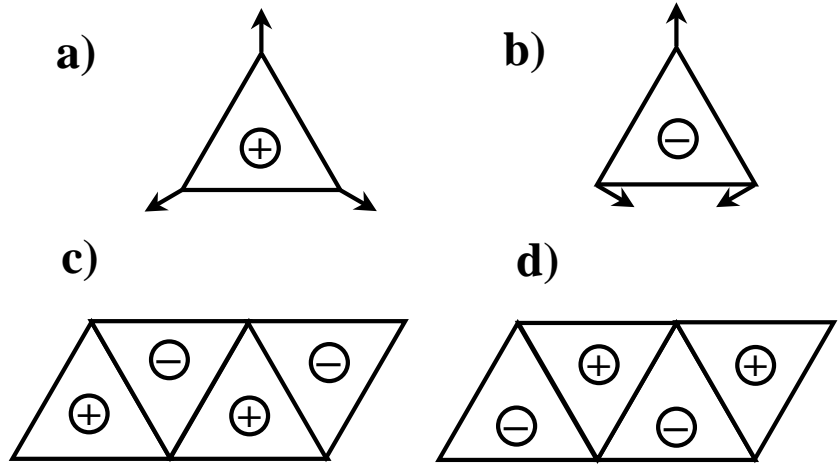}
\caption{
a) and b) $120^\circ$ spin structures of an elementary triangle with positive and negative chirality, respectively; c) and d) Two inequivalent $120^\circ$ spin structures of a triangular antiferromagnet which differ one from another by chiralities of each triangle. They are described by Eq.~(\ref{s}) with the same $\bf a$ and $\bf b$ and with antiferromagnetic vectors ${\bf k}_0$ and $-{\bf k}_0$, respectively. These configurations are referred to in the text as $(+{\bf k}_0)$ and $(-{\bf k}_0)$, respectively.
\label{chir}} 
\end{figure}

In this paper we study chiral fluctuations in triangular antiferromagnets (TAFs) at $T \ll T_N$. The case of TAF in magnetic field ${\bf H}\|\hat c$ ($\hat c$ is a unit vector directed perpendicular to the plain of the lattice) with Dzyaloshinskii-Moriya interaction ${\bf D}\|\hat c$ is considered in detail. This model has been proposed previously to describe quantum TAF CsCuCl$_3$. Expressions for dynamical chirality (or chiral vector) given by Eq.~(\ref{chirality}) describing the chiral fluctuations are derived within the linear spin-wave approximation. We demonstrate that in contrast to non-frustrated systems it is nonzero even at $D,H=0$. As is well known, \cite{mal,mal4} the chiral vector is nonzero only in the case of presence an axial vector (e.g., magnetization of the sample, DM interaction etc.) along which it is directed. We obtain that a ground state of $XY$ and Heisenberg TAFs is characterized by such an axial vector. It is perpendicular to the plane in which spins lie and is directed oppositely for the states with different chiralities of elementary triangles (see Fig.~\ref{chir}). Possibility of investigation of the chiral fluctuations by polarized neutron is also discussed.

\begin{acknowledgments}
I am grateful to S. V. Maleyev for fruitful discussions and for his censorious remarks. It is a pleasure to thank A. G. Yashenkin and D. V. Lebedev for their help in preparation of the manuscript for publication. This work was supported by Russian Science Support Foundation, RFBR (Grant Nos. SS-1671.2003.2, 03-02-17340 and 00-15-96814), Grant Goskontract 40.012.1.1.1149 and Russian Programs "Quantum Macrophysics", "Collective and Quantum Effects in Condensed Matter" and "Neutron Research of Solids".
\end{acknowledgments}

\bibliography{ccc}

\end{document}